# Recycling Silicon Scrap for Spherical Si-C composite as High-Performance Lithium-ion Battery Anodes

*Bhagath Sreenarayanan[a], Marta Vicencio[a], Shuang Bai[b], Bingyu Lu[a], Ou Mao[d], Shiva Adireddy[d], Wurigumula Bao[c,\*], and Ying Shirley Meng[a,b,c,\*]*

[a]Department of NanoEngineering, University of California San Diego, La Jolla, CA 92093

[b]Materials Science and Engineering, University of California San Diego, La Jolla, CA 92093

[c]Pritzker School of Molecular Engineering, University of Chicago, Chicago, IL 60637

[d]ADVANO, Inc. 2045 Lakeshore Drive, New Orleans, LA 70122

*Correspondence to wubao@uchicago.edu; shirleymeng@uchicago.edu

**Keywords:** Lithium-ion Battery; Si-C Composite Anode; Titration Gas Chromatography; Pre-lithiation;

**ABSTRACT**

The growth of the semiconductor and solar industry has been exponential in the last two decades due to the computing and energy demands of the world. Silicon (Si) is one of the main constituents for both sectors and, thus, is used in large quantities. As a result, a lot of Si waste is generated mainly by these two industries. For a sustainable world, the circular economy is the key; thus, the waste produced must be upcycled/recycled/reused to complete the circular chain. Herein, we show that an upcycled/recycled Si can be used with carbon as a composite anode material, with high Si content (~40 wt.%) and loading of 3-4 mAh/cm$^2$ for practical use in lithium-ion batteries. The unique spherical jackfruit-like structure of the Si-C composite can minimize the total lithium inventory loss compared to the conventional Si-C composite and pure Si, resulting in superior electrochemical performance. The superior electrochemical performance of Si-C composites enables the cell energy density of ~325 Wh kg$^{-1}$ (with NMC cathode) and ~260 Wh kg$^{-1}$ (with LFP cathode), respectively. The results demonstrate that Si-based industrial waste can be upcycled for high-performance Li-ion battery anodes through a controllable, scalable, and energy-efficient route.

**Introduction**

The demand for higher energy density lithium-ion batteries (LIBs) has pushed the need to develop higher capacity materials for cathode and anode. On the anode side, alloy-based materials such as Silicon (Si), Germanium (Ge), Tin (Sn), and Aluminum (Al) have been explored[1–3]. The Si delivers a high specific capacity (3579 mAh/g) as the highest among different alloying type anode active material[4], which is quite promising in high energy density LIBs. However, the practical usage of Si anode has been impeded by severe capacity fade due to the large volume change (400%) upon cycling[5]. Several structural design strategies have been proposed to prevent volume expansion, such as Si nanowires via Mg Reduction-Electrospun method[6,7], core-shell models by coating C on Si surface through pyrolysis, carburization methods[8–10] and Si nanolayers with embedded carbon or graphite[11,12]. While these strategies are proven to own high Si content and are effective at the small-scale, large-scale adaptation of these strategies with acceptable cycle stability is still far from reality. This is mainly due to the complex synthesis routes and economics involved in manufacturing, together with the high requirement on performance in practical batteries.

To balance out the specific capacity and cycle stability, several groups have used Si-C-based composite anodes rather than pure Si anodes. Specifically, using amorphous carbon (C) as a component along with Si effectively solves the issues of complex synthesis routes and economics while reducing the capacity fade of Si-based anode. Kwon et al. [13] used a microemulsion method to make a hybrid dual carbon matrix Si-C composite structure using corn starch biomass as a carbon precursor. In this work, the reported volume expansion of the Si-C electrode was 180% after 500 cycles when compared to the pristine state in the half cell. Wang et al. [14] synthesized monodisperse Si and C spheres using magnesiothermic reduction and chemical vapor deposition (CVD), which allowed homogeneous stress-strain distribution during electrochemical cycling. Chae et al. [15] used a process by which petroleum pitch was impregnated inside porous Si to form a porous Si-C nanocomposite. The reported volume expansion was 110% after 49 cycles compared to the pristine state in the half cell. Liu et al. [16] synthesized a hierarchical pomegranate-inspired Si-C structure to tackle the problems of the Si anode. Though most of these works of Si-C in the literature reduce the high stress and strain induced by the volume expansion, they generally focused on the half-cell performance. Also, the structures were synthesized through tedious processes with expensive raw materials.

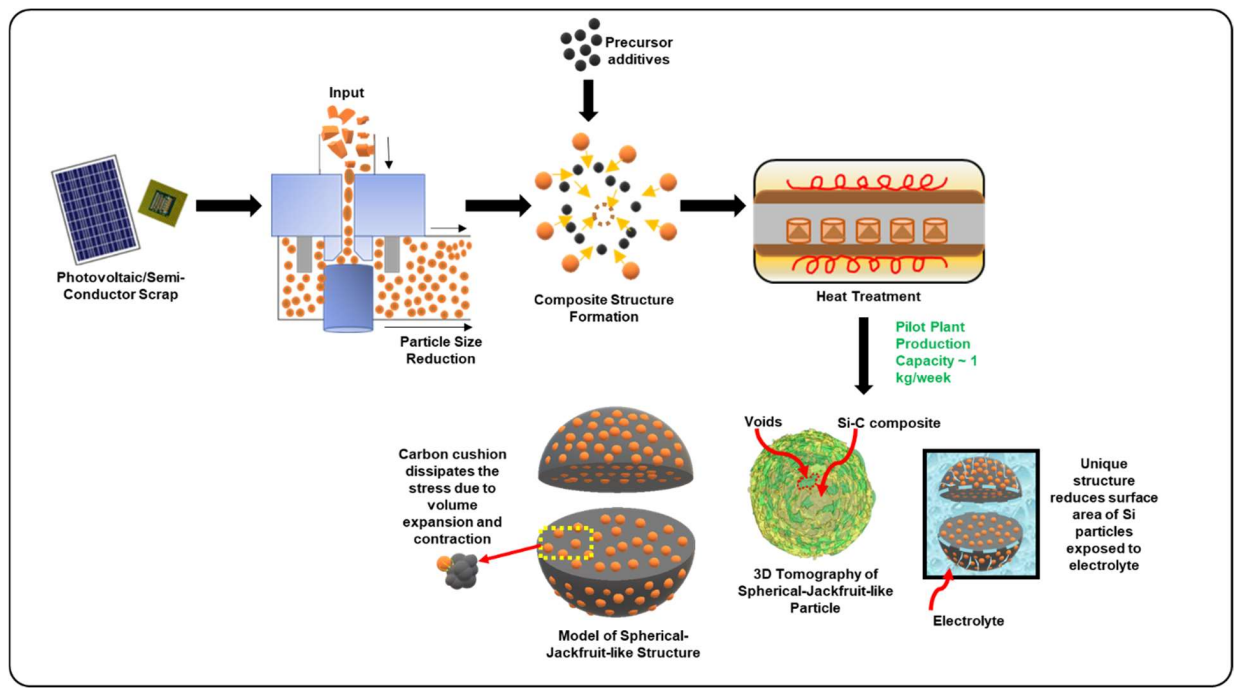

**Figure 1:** Schematic describing the recycling process chain of Si scrap and features of the spherical jackfruit-like structure.

To address the above-mentioned scalability and economy issues of Si based anode materials for LIBs, recycling/upcycling Si materials from other industries should be taken into consideration. In fact, photovoltaic and semiconductor sectors rely heavily on crystalline high-purity Si (~99.9999%); photovoltaic panels have a lifespan of 30 years. By 2050, 80 million tons of solar panels would have reached the end of life[17]. The decommissioned panels, discarded electronic chips, and semiconductor devices contain many toxic and valuable materials, such as high-purity silicon, silver, and copper, which require careful processing before dumping in the landfill. Therefore, innovative recycling and upcycling solutions are necessary to fully harness the value of these materials, thereby enabling a 'circular economy'[17]. This also brings down the future cost of solar panels and semiconductor devices, enabling a transition into a truly renewable economy. In the literature, there have been several works to upcycle silicon scrap to produce value added anode materials. Jin et al[18] used transferred arc thermal plasma method to upcycle silicon scrap collected from photovoltaic cell manufacturing process to produce nano silicon powder to be used as anode for LIBs. Kasukabe et al[19] reported a method by which silicon saw dust waste from silicon ingot preparation is converted into silicon nanoflakes by means of beads milling. The flaky framework was porous in nature to accommodate charge-discharge stress and had excellent cycling stability. Fan et al[20] prepared micro Si@C anodes by recycling lamellar sub-micron silicon from kerf slurry waste using hydrothermal method. Lu et al[21] introduced a method to convert photovoltaic monocrystalline silicon waste into hierarchical silicon/flake graphite/carbon composite. In most of these works in the literature, apart from having complex routes to upcycle the silicon scrap, additional processing steps were used to make complex structures to reduce the problems associated with pure silicon-based anode. Therefore, even though there is a proof-of-concept to upcycle the silicon waste, economically feasible large-scale production (>>1kg/week) of upcycled silicon-based anode is hardly realized.

Herein, we study the spherical Si-C 'jackfruit-like' composite materials, developed through a scalable approach using repurposed Si from semiconductor/solar waste. The detailed process is demonstrated in **Figure 1**. A liquid milling process is first conducted on the Si waste from solar/semiconductor industries to obtain pure Si powder with a designed particle size. The size-reduced Si is then transferred to mix with C precursor and pyrolyzed under inert atmosphere to form the Si-C composite with a spherical jackfruit-like structure. The unique structure of the Si-C composite is proven to benefit cycling performance for the following reasons: 1) the amorphous carbon matrix of the spherical jackfruit-like structure serves as a cushion to withstand volume expansion and contraction cycles of nano Si particles embedded in it. 2) The carbon matrix also serves as a protective shell for the Si particles and exposes less surface area to the electrolyte. 3) The well-connected electronic pathway facilitates electron diffusion through the particles and prevents the formation of trapped Li upon extended cycling[5]. It is found that the as-achieved Si-C composite with 80 nm Si particle size showed a high lithiation capacity of 1400 mAh $g^{-1}$ at 0.14 A $g^{-1}$ with a loading of ~3-4 mAh $cm^{-2}$. By pairing with NMC 622 and LFP cathode, the cells delivered 80% capacity retention after 100 cycles. The Si-C composite from recycled Si shows great potential in commercial applications for next-generation LIBs and can be considered to have a similar cost per kWh as that of commercial graphite anode.

## EXPERIMENTAL METHODS

*Si 150@C and Si 80@C Powder Preparation*

The source of the silicon feedstock used in the production of the Si-C composite is a mixture of photovoltaic and semiconductor-grade silicon scrap. The silicon feedstock vendor (ADVANO Inc) converted this scrap to purified silicon up to 3N purity using an acid treatment process, followed by conversion to micron-sized powder using a dry pulverization process. The manufacturer of the Si-C composite (ADVANO Inc) used a three-step process to convert the feedstock into an Si-C composite. Certain details of the different stages of the process are proprietary to the manufacturer, but general details of the process are provided here. The micron-sized purified silicon feedstock powder is first reduced in size using a wet grinding process (in the presence of organic solvents) to generate a suspension of surface-functionalized silicon nanoparticles. A blend of polymeric carbon precursors and additives is then mixed in the silicon nanoparticle suspension. This suspension of silicon nanoparticles, carbon precursors, and additives is subjected to an atomization process by which an intermediate granulated product of spherical Si-C composite particles are generated. This intermediate product is subjected to a high temperature heat treatment process to reduce the carbon precursor to an amorphous carbon. The final Si-C composite powder is obtained after the end of the heat treatment process.

*Electrode Cast Preparation*

Two types of anode laminates, namely Si 150@C and Si 80@C, were supplied by ADVANO Inc, which consisted of active material (70% by weight), conductive additive (15% by weight), and binder (15% by weight). The 150 and 80 labels represent the primary Si nanoparticle sizes in the composite particle. The active material for this study was made from Si waste supplied by a producer of metallurgical grade silicon. A proprietary ADVANO formulation and method were utilized to prepare the active material of the spherical Si-C jackfruit-like composite. For the control experiment, electrode cast consisting of primary nano Si and carbon mixture—obtained after milling the powder with spherical jackfruit-like structure in the Thinky Mixer (particle size of primary nano Si is 150 nm (ADVANO Inc), active material—70% by weight), Super C-65 (MTI) as conductive additive (15% by weight) and Polyacrylic Acid ( Mv 450,000, Sigma Aldrich) as binder (15% by weight) was prepared. The electrode cast mixture was dispersed in N-Methyl-2-Pyrrolidone (NMP) and mixed using a Thinky Mixer at 2000 rpm for 1 hour. The obtained slurry was cast onto a copper foil using the Doctor blade and was dried for 10 h at 120°C under vacuum to remove the NMP.

*Electrochemical Testing*

For the half-cell testing, a working electrode (9.5 mm in diameter) was assembled into a 2032 type coin cell in an Ar-filled glove box. Li metal (1 mm thick) was employed as the counter electrode. The electrolyte was 1.2 M LiPF$_6$ in Ethylene Carbonate (EC): Ethyl Methyl Carbonate (EMC)=3:7 (wt.%) +10 wt.% Fluoroethylene Carbonate (FEC)+1 wt.% Lithium difluorophosphate (LiPO$_2$F$_2$). The anode half-cell was discharged at 0.1 C to 10mV, with constant voltage at 10mV until C/20 and then charged at 0.1 C to 1.5 V (for the first 2 formation cycles). For subsequent cycles, the anode half-cell was discharged at 0.3 C to 10mV, with Constant Voltage at 10mV until C/20 and then charged at 0.3 C to 1.5 V. For the full cell testing, the negative electrode was paired with LiNi$_{0.6}$Mn$_{0.2}$Co$_{0.2}$O$_2$ (NMC622) cathode (areal loading ~3.7 mAh/cm$^2$) and LiFePO$_4$ (LFP) cathode (areal loading ~3 mAh/cm$^2$) (NMC622 from Targray, LFP from NEI, 90 wt.% active material, with 5 wt.% C-65 carbon as conductive agent, and 5 wt.% PVDF as binder) and assembled in a 2032 type coin cell in Ar-filled glovebox. For NMC622 cathode testing, the half cell was charged to 4.3 V at C/10 and then discharged to 3 V at C/10 for 2 cycles. For the subsequent cycles, the cell was charged to 4.3 V at 0.3 C and then discharged to 3 V at 0.3 C. For LFP cathode testing, LFP half-cell was charged to 3.6 V at C/10 and then discharged to 2 V at C/10 for 2 cycles. For the subsequent cycles, LFP half cell was charged to 3.6 V at 0.3 C and then discharged to 2 V at 0.3 C. The NMC622 full-cell was charged at 0.1 C to 4.2 V, with Constant

Voltage at 4.2 V until C/40 and then discharged at 0.1 C to 2.7 V (for the first 2 formation cycles). For subsequent cycles, the full cell was charged at 0.3 C to 4.2 V, with Constant Voltage at 4.2 V until C/40 and then discharged at 0.3 C to 3 V. The LFP full cell was cycled between 2 V and 3.6 V at C/10 for the first two cycles and C/3 for subsequent cycles. The N/P ratio used in the NMC622 full cells was 1.1-1.2 whereas for LFP full cells was 1.3-1.4.

*Pre-lithiation*

For pre-lithiation, the mechanical shorting method was employed. In this method, a few drops of electrolyte were added on the surface of the punched anodes. Li metal chip was placed on top of the electrode and a 200 g 'weight' was placed on top of the Li metal chip. This setup was kept idle for 10 minutes, after which the pre-lithiated electrode was obtained and washed with the Dimethyl Carbonate solvent to remove excess salt and solvent before assembling the pre-lithiated full cell.

*Characterization*

**Titration Gas Chromatography (TGC)**

The TGC experiments were performed using a Shimadzu GC-2010 Plus Tracera equipped with a barrier ionization discharge (BID) detector. The Split temperature was kept at 200 °C with a split ratio of 2.5 (split vent flow: 20.58 ml·min$^{-1}$, column gas flow: 8.22 ml·min$^{-1}$, purge flow: 0.5 ml·min$^{-1}$). Column temperature (RT-Msieve 5A, 0.53 mm) was kept at 40 °C, and the BID detector was held at 235 °C. Helium (99.9999%) was used as the carrier gas, and the BID detector gas flow rate was 50 ml·min$^{-1}$. The electrode sample was put in a septum sealed glass vial, and after injecting the 0.5mL acetic acid/ethanol (200 proof anhydrous), the sample gases (30 μL) were injected into the machine via a 50 μL Gastight Hamilton syringe.

**Scanning Electron Microscopy (SEM)**

The SEM was conducted on the FEI Apreo SEM; the coin cells were disassembled in the Ar-filled glovebox after cycling. The samples were transferred to the SEM chamber for cross-section analysis and surface analysis with minimal exposure to air. The electron beam operating voltage was 5 kV and the operating current was 0.1 nA.

**Cryogenic Dual Beam Focused Ion Scanning Electron Microscopy (Cryo FIB-SEM)**

The FIB-SEM was conducted on the FEI Scios Dual-beam microscopy; the samples were disassembled in the Ar-filled glovebox after cycling. The samples were transferred to the FIB chamber without any exposure to air. The electron beam operating voltage was 5 kV, and the stage was cooled with liquid nitrogen to −180 °C. Sample cross-sections were exposed using a 1 nA ion beam current and cleaned at 0.1 nA.

**X-Ray Diffraction (XRD)**

To investigate the crystal structure of the composite particles, X-Ray Diffraction (XRD) was conducted with a Rigaku Smart Lab Diffractometer. The scan speed was 2 s and the scan step was 0.02 deg with Braggs-Brentano Focusing Mode. The μSi and Nano Si powders (Alfa Aesar) were used to generate baseline XRD pattern.

**Electrochemical Impedance Spectroscopy (EIS)**

The EIS measurements were performed from 1 MHz to 10 mHz with an applied potential of 10 mV using Bio-Logic VSP 150 at 25°C. The results were fitted using Z-View Software.

**Thermogravimetric Analysis and Differential Scanning Calorimetry (TGA/DSC)**

The TGA/DSC Analysis was performed using SDT 650 machine. The temperature increased from 50°C-1000°C at the rate of 10°C/min in air.

**Raman Spectroscopy**

The Raman Spectroscopy measurements were performed using Renishaw inVia Raman Microscope. The measurements were run using a 532 nm green laser source with 1800 L mm$^{-1}$ grating and with 20x magnification. The μSi, Graphite and Hard Carbon powders were used to generate baseline Raman Spectra.

**RESULTS AND DISCUSSION**

**Structure, Morphology, and Composition of Si 150@C and Si 80@C composite**

The structural morphology of the two types of Si-C composite secondary particles (Si 150@C and Si 80@C) were characterized using SEM (**Figure S1 (a and b)**) and FIB-SEM (**Figure 2 (d and e)**). From the SEM images, the secondary particle distribution is similar for Si 150@C and Si 80@C powders, with larger secondary particles being 10-13 μm and a higher percentage of particles <2 μm. However, from FIB-SEM images, it can be confirmed that Si 150@C has a larger primary particle size (primary particle~150 nm) when compared to Si 80@C (primary particle size ~80 nm). To quantify the porosity of the uniquely designed structure, 3D reconstruction was done to obtain the 3D Tomography model (**Supplementary Video 1**). The green color in the 3D Tomography image indicates the 'pores' in the Si 150@C secondary particle whereas yellow color indicates the Silicon-Carbon composite material. The porosity of Si 150@C was determined to be 18.4%. Further, to determine the type of carbon in these composites, Raman measurement was performed for Si 150@C powder and compared with the Raman Spectra obtained for hard carbon, graphite, and commercial micro-Si powder (**Figure S1 (c)**). From the Raman Spectra, Si 150@C showed the crystalline Si peak at 520 cm$^{-1}$ and Si-OH peak at 947 cm$^{-1}$, consistent with the Raman spectra of commercial μSi powder. Also, between 1300 cm$^{-1}$ and 1600 cm$^{-1}$, the Raman spectra of Si 150@C were similar to that of hard carbon, indicating that the composite structure has hard carbon with an $I_D/I_G \sim 0.994$. The composition of the secondary particle was analyzed using TGA/DSC (**Figure 2 (a and b)**). From the compositional analysis using TGA/DSC, there was a weight loss of 47.78% and 47.17% at 600°C for Si 150@C and Si 80@C, respectively, which corresponded to the carbon content. This corresponds to the formation of $CO_2$, which corresponds to the carbon content of the Si-C composite. Further, there was a formation of $SiO_2$ beyond 1000°C, and the increase in weight was 46.75% and 33.70%, respectively, for Si 150@C and Si 80@C. This weight increase corresponds to the Si content of the Si 150@C and Si 80@C. The Oxygen amount of the composite structure was also quantified using TGA/DSC as explained in **Note S1**. These results show that Si 150@C has a higher Si (46.75%) and lower Oxygen content (6%) whereas Si 80@C has a lower Si (33.7%) and higher Oxygen content (12%). The corresponding EDS mapping of the individual Si-C composite secondary particle from two samples (**Figure S2 (a and b)**) shows homogeneous distribution of Si and Carbon in the composite structure.

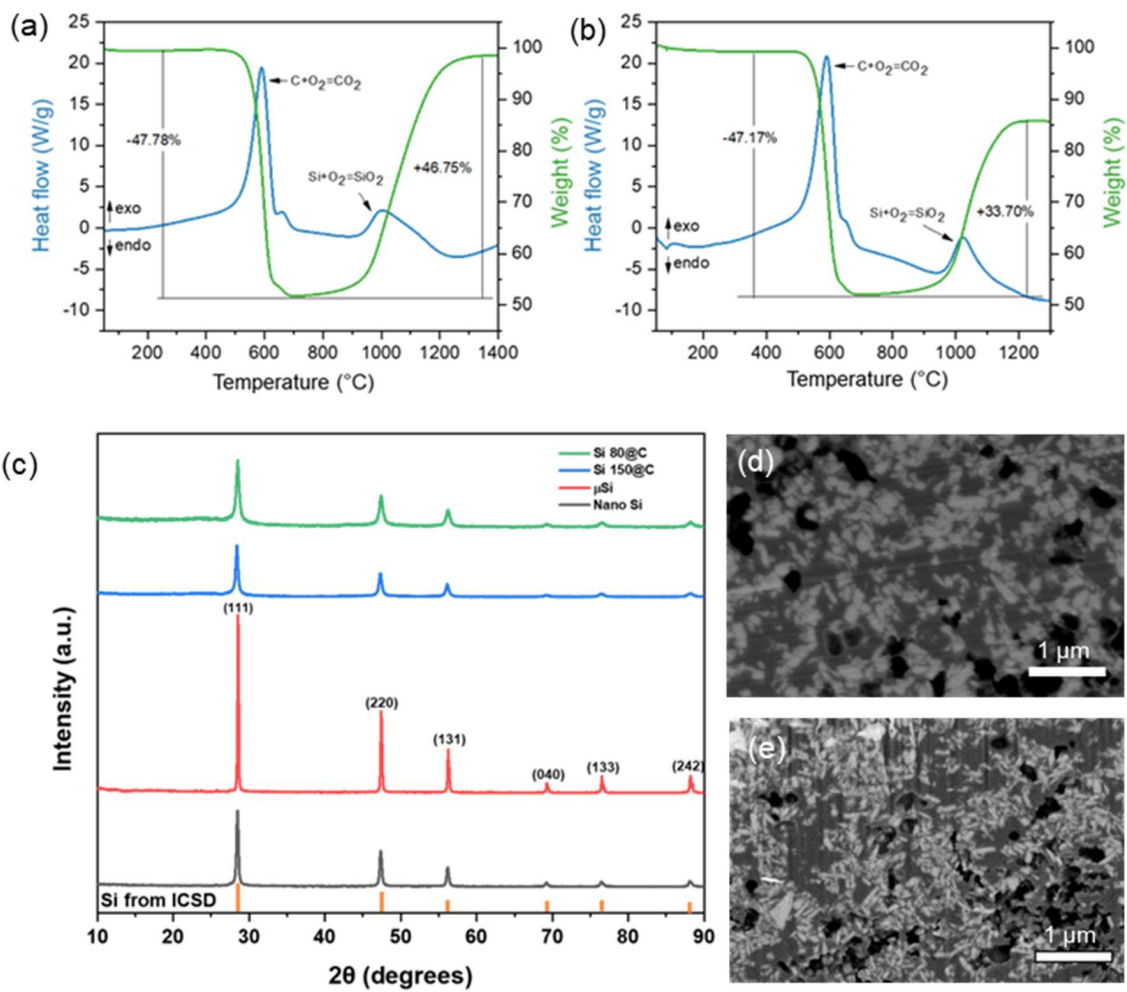

**Figure 2** TGA/DSC of (a) Si 150@C and (b) Si 80@C. XRD of (c) Si 150@C and Si 80@C with μSi and nano Si as reference. Cryo-FIB-SEM of cross-section of secondary particle of (d) Si 150@C and (e) Si 80@C

The XRD was performed on Si 150@C and Si 80@C to determine the crystal structure of the particles. **Figure 2 (c)** shows that both the composite materials show crystalline Si signals as observed in μSi and nano Si. There were no carbon/graphitic signals observed in the XRD pattern indicating that carbon in the composite was amorphous in nature.

**Electrochemical Performance of Si 150@C and Si 80@C composite**

A comparative study on electrochemical performance with pure nano Si (Nano Si (150)), nano Si/carbon mixture (Si/C), Si@150, and Si@80 were conducted to understand the morphology influence. From **Figure 3 (a)**, it can be seen that the initial Columbic Efficiencies (ICEs) of Nano Si (150) and Si/C were 73.34% and 75.52%, respectively. In contrast, the ICEs of Si 150@C and Si 80@C were 86.85% and 83.37%, respectively. The data indicate that the spherical jackfruit-like secondary particle structure can lower the initial capacity losses in the electrode level so that the ICE can be improved. Both samples gave an initial lithiation capacity more than 1400 mAh g$^{-1}$ with an areal loading of ~3-4 mAh cm$^{-2}$, while the Si 80@C had slightly lower capacity (1400 mAh g$^{-1}$) than Si 150 @C (1600 mAh g$^{-1}$). The reduction in ICE and lithiation capacity were possibly due to the slightly increase in oxygen content and the overall surface area increase of the nano Si particles when the particle size was smaller. From the cycle performance shown in **Figure 3 (b)**, the capacity retention for Nano Si (150) and Si/C were 56% and 33.85%, with the corresponded average CEs 97.81% and 96.34%, respectively. For Si 150@C, the capacity retention was 75.79% after 20 cycles, and the average CE was 98.47%. This data further confirms the integrity of the spherical jackfruit-like structure in reducing capacity loss. Si/C composite is a mixture of Si and C powder which does not have a spherical jackfruit-like structure. Therefore, the absence of carbon cushion leads to exposure of silicon particles to the electrolyte compared to the Si 150@C and Si 80@C, resulting in more active Li loss due to SEI formation. Additionally, severe volume expansion can cause delamination and loss of electronic pathways, leading to trapped Li. In contrast, a spherical jackfruit-like structure provides a carbon cushion that dissipates stress due to volume expansion and builds a 3D electronic pathway, ensuring good cyclability. When the primary particle size was reduced to 80 nm, in the case of Si 80@C, the capacity retention was 90.42%, and the average CE was 99.11%. The improved capacity retention for Si 80@C, when compared to Si 150@C, was possibly due to the reduction in primary particle size and the increase in oxygen content in the Si-C composite structure, which plays a crucial role in stabilizing the interphase[22]. Our previous work also shown that a reduction in particle size could reduce the trapped Li[5]. It should also be noted that in the **Figure 3 (b)**, for Si-C composite without the spherical jackfruit like structure and also for Nano Si(150), there is a rapid capacity fade due to the fact that the silicon particles are not protected by the carbon cushion anymore and are exposed to electrolyte.

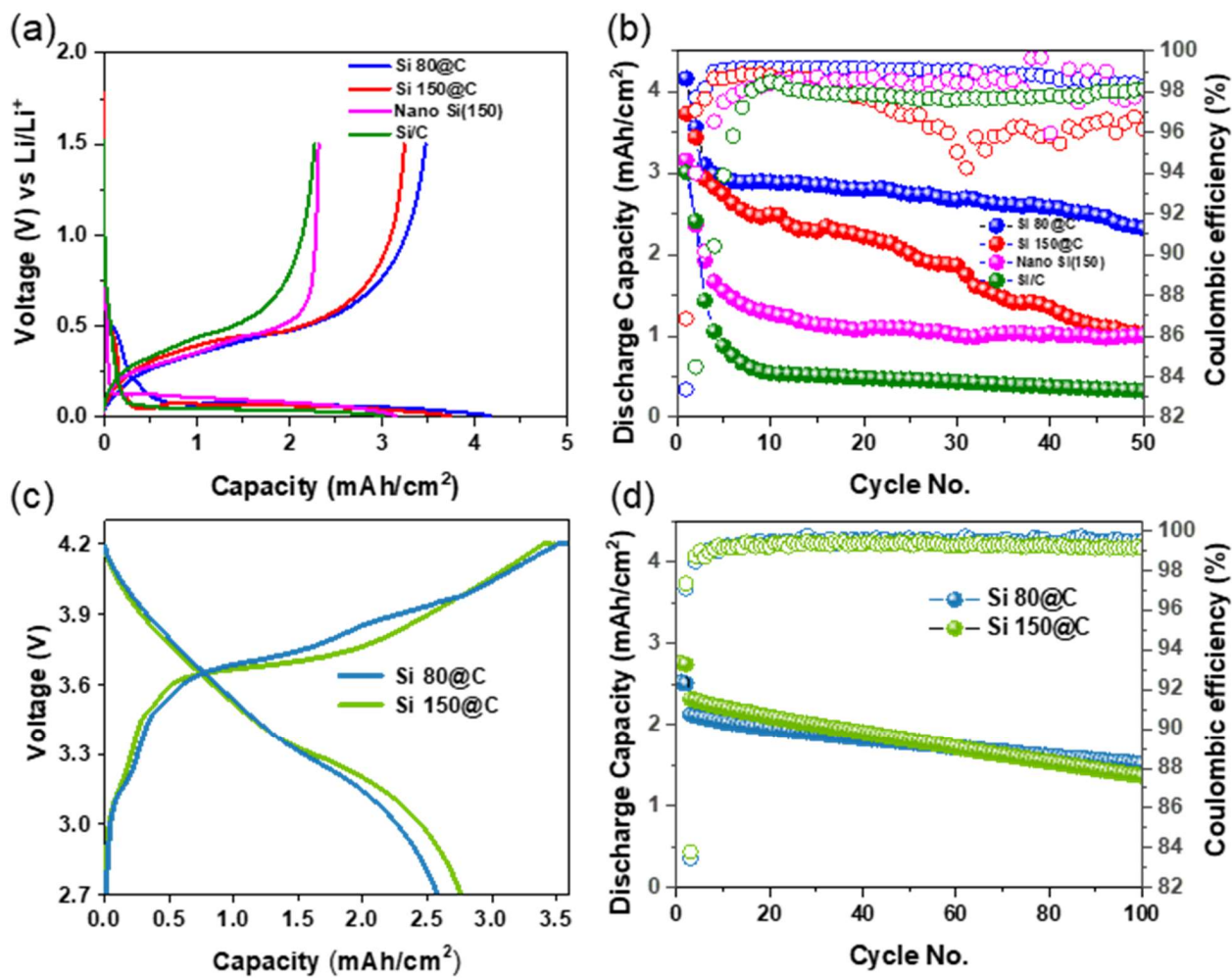

**Figure 3** (a) Charge-Discharge profile of Nano Si (150), Si/C, Si 150@C, Si 80@C half-cell at first cycle and (b) following cycle performance (c) Charge-Discharge profile of Si 150@C-NMC622 and Si 80@C-NMC622 full cells at first cycle (d) following cycling performance.

The full cell was made using Si 150@C and Si 80@C as the negative electrode and NMC622 as the positive electrode. The first cycle voltage profile and subsequent electrochemical cycling performance of NMC622 is shown in **Figure S3 (a and b)**. The designed N/P ratio for the full cell was 1.1-1.2. From **Figure 3 (c)**, the ICE for Si 150@C and Si 80@C full cells with NMC622 cathode were 79.13% and 70.3% respectively. The lower ICE for Si 80@C full cell compared to Si 150@C full cell was consistent with the observation in half cell. From **Figure 3 (d)**, it can be observed that the capacity retention for Si 150@C and Si 80@C full cells were 58.95% and 71.22% and the average CE was 99.11% and 99.29%, respectively, after 100 cycles which were in good agreement with the half-cell data. The results presented in **Figure 3 (b and d)** demonstrate that the Si 150@C and Si 80@C exhibit faster half-cell decay rates compared to full cells. In our previous research on micro silicon (μSi) anodes[5], it was noted that reducing the region of volume expansion by controlling the state of charge can mitigate total lithium inventory losses in silicon-based anodes. To achieve this in full cells, we increased the Negative to Positive (N/P) ratio which in this case is 1.1-1.2. However, in the half cells investigated in this study, we discharged them up to full capacity utilization of the anode (constant voltage of 10 mV), which caused maximum volume expansion in each cycle and led to poor cyclability, similar to running a full cell with a very low N/P ratio. It must be noted that pre-lithiation strategy is quite necessary for silicon-based anode to substitute for lost Li in the first cycle.[23–25]

**Quantification of Li inventory loss in Half Cell**

To study how the Li inventory losses evolve in the Si 80@C and Si 150@C, the Titration Gas Chromatography (TGC) method was employed in half cell system. The Li in the cycled anode is present as trapped Li and SEI (Li$^+$), in which only trapped Li can react with the protic solvent to generate the H$_2$, and GC can quantify the amount. Our previous work established the TGC method for pure Si anode using ethanol as the titration solvent because Si is not stable in base solution and water[5]. Sulfuric acid is necessary for quantifying the trapped Li in the carbon-based anode[26]. However, sulfuric acid ($K_\alpha = 1 \times 10^3$) has higher acidity than HF ($K_\alpha = 6.6 \times 10^{-4}$), therefore, it can react with LiF - one of the SEI components to produce HF, which reacts with Si to generate additional $H_2$. To accurately quantify the trapped Li in Si-C composite, the titration solvent should achieve the following requirements: 1) No base solution or pure water 2) Would not generate HF with SEI components, which indicates that the acidity of the solvent should be lower than HF. Acetic acid with the acidity of ($K_\alpha = 1.8 \times 10^{-5}$) meets all the requirements. To verify the chemical stability of the titration solvent with SEI components and Si-C composite, a comparative study was done with sulfuric acid and acetic acid as the titration solvent. The solvents are added to the mixture of the pristine Si-C sample and LiF, the results are shown in **Figure S4 (a)**. It was found that sulfuric acid produced hydrogen gas upon reacting with LiF and the Si-C composite. In contrast, no hydrogen gas was produced with acetic acid as the titration solvent, indicating that acetic acid is stable with LiF. Therefore, acetic acid was chosen as the titration solvent for the Si-C composite electrode. For developing the calibration curve, a control experiment was performed where four identical Si 150@C half cells were de-lithiated to four different points of de-lithiation (PD-X, where X=1,2,3,4) corresponding to 0.5 mAh, 1 mAh, 1.4 mAh, 2 mAh respectively, as shown in **Figure S4 (b)**. The four cells were disassembled, and the Si 150@C electrodes were transferred to penicillin bottles, to which acetic acid was added to liberate $H_2$ gas. The liberated $H_2$ gas was quantified, and the calibration curve was plotted, as shown in **Figure S4 (c)**. After the development of the calibration curve, the TGC test was conducted on Si 150@C and Si 80@C half cells after 2 formation cycles, 10 cycles, 30 cycles, and 55 cycles to understand the evolution of lithium inventory losses in the half-cell system (**Figure 4 (a) and (b)**). For the case of Si 150@C

half cells, the trapped Li increased from 0.14 mAh cm$^{-2}$ after 2 cycles to 1.379 mAh cm$^{-2}$ after 55 cycles, and SEI increased from 0.52 mAh cm$^{-2}$ after 2 cycles to 1.579 mAh cm$^{-2}$ after 55 cycles. However, for the case of Si 80@C half cells, the trapped Li increased from 0.133 mAh cm$^{-2}$ after 2 cycles to 0.573 mAh cm$^{-2}$ after 55 cycles, and SEI increased from 0.666 mAh cm$^{-2}$ after 2 cycles to 1.715 mAh cm$^{-2}$ after 55 cycles. From the data, it can be concluded that when the primary particle is reduced inside the secondary particle of Si-C composite, the total Li inventory losses are reduced. Also, there was a significant reduction in the trapped Li.

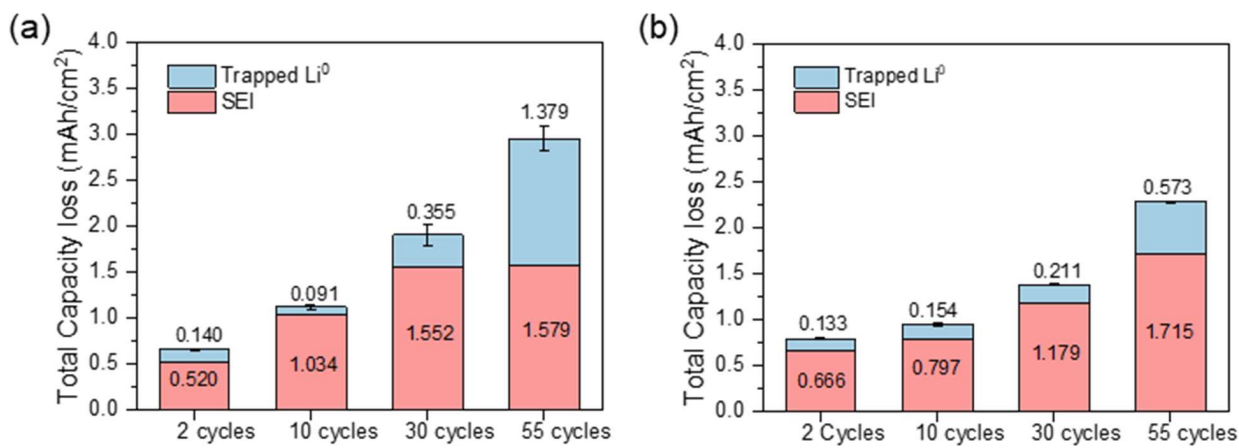

**Figure 4** TGC Data of (a) Si 150@C half cells (b) Si 80@C half cells.

**Post Mortem Analysis of Composite Full Cell**

To further understand the differences of the two types of Si-C composite in a full cell with limited Li resource for practical applications, Electrochemical Impedance Spectroscopy (EIS) and Cryo-FIB-SEM were performed after several charge-discharge cycles. **Figure 5 (a)** shows the EIS spectra of Si 150@C full cell after different cycle numbers, and **Figure 5 (b)** shows the corresponding values of equivalent series resistance ($R_s$) and various Charge Transfer Resistances ($R_{ct}$) obtained from the fitted circuit. The $R_s$ is generally influenced by components in the cell setup and the value of $R_s$ relatively remains constant as the full cell is cycled indicating the cells shows good consistency[27]. The various $R_{ct}$ in the EIS spectra are a combination of different resistances, i.e., SEI and CEI resistances, interphase resistance between carbon matrix and nano Si in the jackfruit-like structure. The various Constant Phase Elements (CPEs) are the non-ideal capacitive behaviour between different interphases such as nano Si and carbon matrix, nano Si and electrolyte, and carbon matrix and electrolyte. For Si C@150 full cell, the $R_{ct}$ increases with cycle number from 1.721 $\Omega$ in 3$^{rd}$ cycle to 12.446 $\Omega$ in the 50$^{th}$ cycle. Also, it can be seen that after the 3rd cycle, additional CPEs and resistances are required to fit the EIS spectra, indicating the possible formation of new interphases that affect the overall charge transfer process. On the contrary, for Si 80@C full cell (**Figure 5 (d) and (e)**), a higher value of $R_{ct}$ was observed in the case of Si 80@C compared to the Si 150@C at the beginning which is possibly due to the higher surface area between Si and carbon due to the smaller particle size. However, a lower $R_{ct}$ increment from 7.305 $\Omega$ in 3$^{rd}$ cycle to 8.202 $\Omega$ in 50$^{th}$ cycle is observed in Si 80@C full cell, indicating the morphology of the anode preserved well upon cycling.

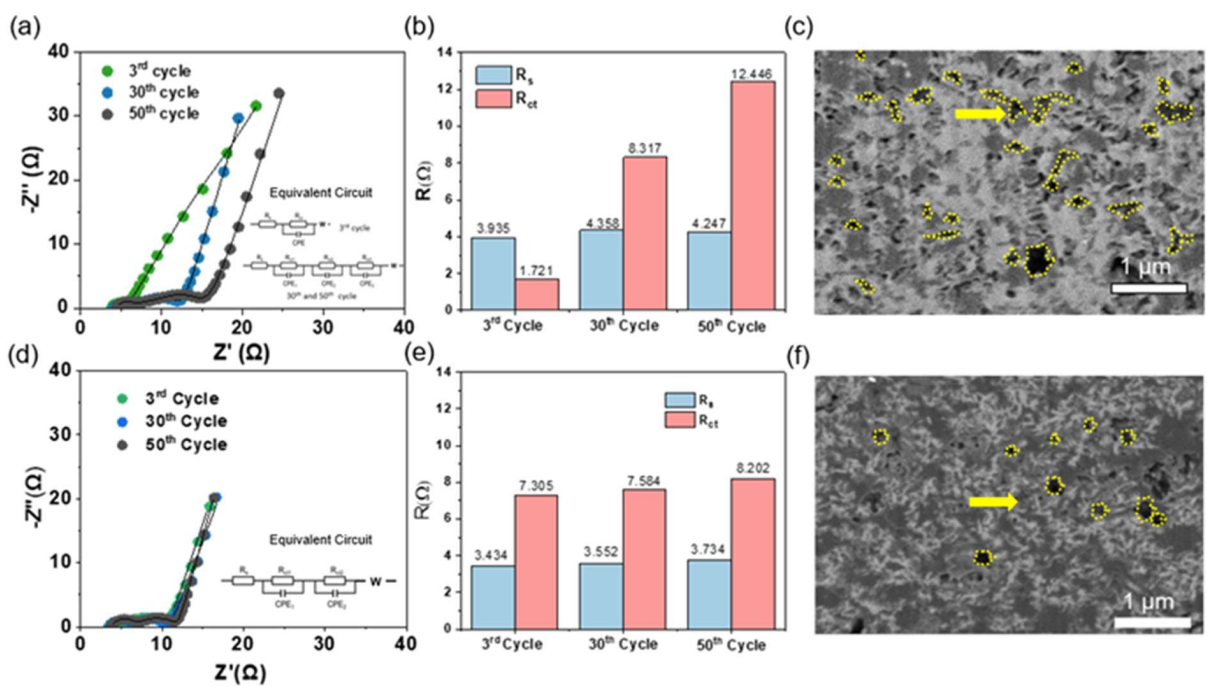

**Figure 5** (a) EIS results of Si 150@C-NMC 622 full cell at 3$^{rd}$, 30$^{th}$ and 50$^{th}$ cycle and (b) related impedance value from equivalent circuit, (c) Cryo-FIB-SEM of cross-section of Si 150@C secondary particle after 50$^{th}$ cycle. (d) EIS results of Si 80@C-NMC 622 full cell at 3$^{rd}$, 30$^{th}$ and 50$^{th}$ cycle and (e) related impedance value from equivalent circuit, (f) Cryo-FIB-SEM of cross-section of Si 80@C secondary particle after 50$^{th}$ cycle.

To understand the morphological evolution of the Si-C composite particles upon cycling, the secondary particles of Si 150@C and Si 80@C were milled using Cryo-FIB. The milled sections were imaged using SEM with backscatter mode as shown in **Figure 5 (c and f)** after 50 cycles. The 'black' spots (outlined with yellow color) on the SEM images indicate voids possibly formed after cycling. It is clear that Si 80@C has less voids formed than the Si 150@C which indicates Si 150@C is more susceptible to pulverization than Si 80@C. It was also in good agreement with the TGC data that Si 150@C sample shows a higher chance of losing the electronic pathway, resulting in more trapped Li formation, compared to Si 80@C. At the electrode level, it can be seen that Si 150@C electrode undergoes volume expansion of about ~97% after 50 cycles compared to the pristine electrode (**Figure S5 (a-c)**). However, Si 80@C electrode undergoes volume expansion of about ~55% after 50 cycles (**Figure S5 (d-f)**), which is consistent with the EIS results that the full cell with Si 150@C electrode show higher interface impedance than Si 80@C electrode.

To summarize the morphological properties, in the full cell system, cell impedance grows steadily for Si 150@C-NMC622 full cell after 3 cycles indicating the formation of new interfaces, which is confirmed by the higher percentage of cracks seen in **Figure 5 (c)**. For the case of Si 80@C-NMC622 full cell, the cell impedance grows sparingly after 3 cycles due to the lesser number of interfaces formed, which is consistent with the less cracks seen in **Figure 5 (f)**. The appearance of a higher percentage of cracks in Si 150@C electrode in the full cell is further supported by the higher percentage of volume expansion at the electrode level when compared with the Si 80@C electrode. As far as SEI composition is concerned on the anode side, FEC as an additive decompose prior to the decomposition of EC and EMC owing to its high reduction potential due to the fluorination, forming polymeric SEI[28]. Thus, it can reduce the decomposition of EC and EMC and also lead to reduced consumption of the salt. The SEI primarily consists of the following functionalities: aliphatic carbons (C ($sp^3$) based groups), carboxyl groups (ROCO), alkoxy groups (RCO), ionic carbonate salts ($RCO_3^-$), inorganic lithium compounds such as $Li_2CO_3$. LiF, $Li_2O$, fluorophosphoro oxides ($P_xO_yF_z$)—which could come from the decomposition of $LiPO_2F_2$ salt additive as well[29]. Due to the presence of significant amount of oxygen in the composites for both Si 150@C and Si 80@C, glassy SEI products such as $Li_xSi_yO_z$ can also exist which is generally reported for $SiO_x$ anodes[30–32]. These glassy products are generally irreversible upon formation and have low electronic and high ionic conductivities[33]. The irreversibility of products such as $Li_xSi_yO_z$ leads to low Coulombic Efficiency in the first formation cycle in these composites—consistent with our half-cell and full cell data. The surface oxide layer can also lead to the formation of surface etched products such as $SiF_x$ in the presence of fluorine containing additives such as FEC and $LiPO_2F_2$[29]. As far as effect of $LiPO_2F_2$ is concerned, it is generally used as an additive to stabilize the cycled cathode especially the high Nickel layered oxide cathodes[34,35]. However, it has also been reported that $LiPO_2F_2$ can have a positive effect on cycling of silicon anode by formation of $-PO_2F_2$ ions which can continuously protect the silicon surface upon cycling[36]

**Pre-lithiation strategy to reduce the first cycle loss in full cell for practical applications**

Considering the insufficient ICE when using Si based anode, the pre-lithiation strategy was applied on the Si 80@C electrode and paired with LFP cathode for demonstration. The LFP was chosen as the cathode because previous reports have suggested overcharging issues (highly de-lithiated states) in NMC based cathodes in practical applications[37], thereby not giving a true picture of Si 80@C electrode's performance in full cell for practical application. Also, LFP is the most widely used Co-free Li ion battery cathode in the market. The first cycle voltage profile and subsequent electrochemical cycling performance of LFP is shown in **Figure S3 (c and d).** The full cell was designed for N/P ratio = 1.4 before pre-lithiating the Si 80@C electrode. Also, a reference full cell with LFP cathode was assembled with N/P ratio = 1.4 without pre-lithiation. From **Figure 6 (a)**, the ICE of Si 80@C-LFP full cell was 75.53%. However, the pre-lithiated Si 80@C electrode (Preli-Si 80@C) for 10 minutes, the ICE of the full cell increased to 85.71%. From **Figure 6 (b)**, it can be seen that the capacity retention of Si 80@C-LFP full cell was 56.14% after 100 cycles. In contrast, pre-lithiated Si 80@C-LFP full cell had a capacity retention of 80.68% after 100 cycles. Also, the average CE was 99.86% for the pre-lithiated Si 80@C-LFP full cell and 99.29% for the Si 80@C-LFP full cell after 100 cycles. Upon carefully analyzing the Si-C published works in the literature, it can be concluded that most of the full cell data reported has the maximum areal capacity (charge/discharge) < 2.5 $mAh/cm^2$ and silicon content < 50% in the anode (**Figure S6**

**(a)**). The **Figure S6 (b)** suggests that long cycling in the full cell (>200 cycles) comes at the cost of compromising on areal capacity or silicon content in the anode. Therefore, for evaluating the practicality of a new Si-C anode technology, the following are the critical metrics to consider: Si content in the anode (%), maximum areal capacity achievable, capacity retention, and cycle number. To our knowledge, ~2.75 mAh/cm² Si-C based anode full cell with a silicon content of ~30-35 wt%, with cycle retention of ~80% after 100 cycles, is hardly reported. The superior electrochemical performance of Si-C composites enables practical pouch cell energy density of ~325 Wh kg$^{-1}$ with NMC cathode and Si 150@C anode (**Table S1**) and ~260 Wh kg$^{-1}$ with LFP cathode and Si 80@C anode (**Table S2**), respectively.

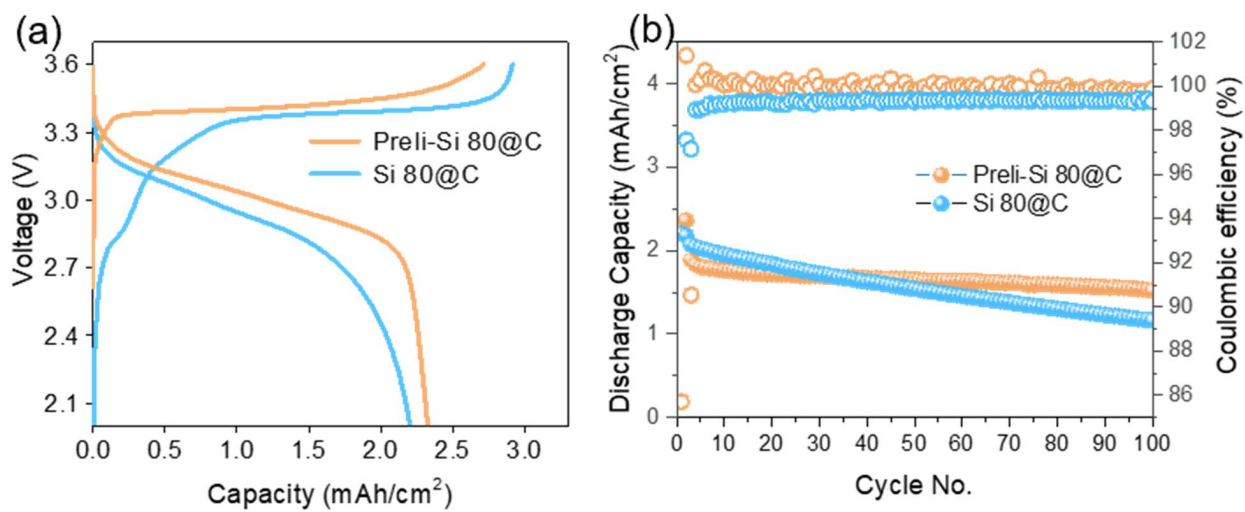

**Figure 6** (a) Charge-Discharge profile of Si 80@C-LFP and Preli-Si 80@C-LFP full cells at first cycle (b) following cycling performance

**CONCLUSIONS**

In summary, a scalable industrial process was utilized for recycling the Si from photovoltaic and semiconductor waste to develop a unique lithium-ion battery Si-C anode material with a spherical jackfruit-like structure. The optimized Si-C composite can deliver a capacity of 1400-1600 mAh/g at a cost expected to be similar to that of Graphite. The unique spherical jackfruit-like structure embeds the crystalline nano Si particles inside the amorphous carbon matrix. The designed structure reduces the surface area of materials exposed to the electrolyte and maintains a good electronic pathway. In addition, the carbon matrix acts as a cushion to dissipate the stress due to the Si volume expansion and contraction upon cycling. The spherical jackfruit-like nature of the material allows the packing of high Si content in limited volume without compromising on structural integrity. The Si-C composite anode shows improved cycling performance compared to the nano Si anode. Besides, the Si-C composite anode with a smaller particle size (Si 80@C) had higher capacity retention and average coulombic efficiency upon cycling compared to the Si-C composite anode with a larger primary particle size (Si 150@C). The TGC and the EIS results show that the improved cycling stability in Si 80@C could be attributed to the reduced lithium inventory loss and lower impedance growth resulting from small Si particle size and improved stress dissipation. The designed Si 80@C composite anode is compatible with NMC 622 and LFP cathodes. With the prelithiation strategy, the Si 80@C- LFP full cell shows 80% capacity retention after 100 cycles with practical cathode loading of ~3 mAh/cm$^2$. The results demonstrate the promising upcycling avenue for producing low-cost and high-performance Si-based anodes for next-generation lithium-ion battery systems.

## AUTHOR INFORMATION

**Corresponding Author**

*(WB) E-Mail: wubao@uchicago.edu

*(YSM.) E-mail: shirleymeng@uchicago.edu

**ORCID:**

Wurigumula Bao: 0000-0001-8109-1546

Ying Shirley Meng: 0000-0001-8936-8845

**Notes**

The authors declare no conflicts of interest.

ACKNOWLEDGEMENTS

The authors gratefully acknowledge funding supported by the ADVANO Inc. Ying Shirley Meng is the scientific advisor for ADVANO Inc. The SEM and Cryo-FIB-SEM was performed at the San Diego Nanotechnology Infrastructure (SDNI) of UCSD, a member of the National Nanotechnology Coordinated Infrastructure supported by the National Science Foundation (Grant ECCS-1542148). The electrochemical testing was performed on a battery test station donated by NeWare.

AUTHOR CONTRIBUTIONS

B.S., W.B., and Y.S.M. conceived the ideas. B.S. and W.B performed TGC measurements. W.B. and B.L performed TGA/DSC, Cryo-FIB-SEM experiments. B.S and M.V. conducted electrochemical experiments. B.S. completed cross-section SEM. S.B. O. M. and S.A. was involved in the technical discussion. All authors discussed the results and commented on the manuscript.

REFERENCES

1. Li, H. *et al.* Circumventing huge volume strain in alloy anodes of lithium batteries. *Nat Commun* **11**, 1584 (2020).
2. Zhang, W.-J. A review of the electrochemical performance of alloy anodes for lithium-ion batteries. *J Power Sources* **196**, 13–24 (2011).
3. Wang, H. *et al.* The progress on aluminum-based anode materials for lithium-ion batteries. *J Mater Chem A Mater* **8**, 25649–25662 (2020).

4. Zhang, X. *et al.* Geometric design of micron-sized crystalline silicon anodes through in situ observation of deformation and fracture behaviors. *J Mater Chem A Mater* **5**, (2017).

5. Sreenarayanan, B. *et al.* Quantification of lithium inventory loss in micro silicon anode via titration-gas chromatography. *J Power Sources* **531**, 231327 (2022).

6. Favors, Z. *et al.* Towards Scalable Binderless Electrodes: Carbon Coated Silicon Nanofiber Paper via Mg Reduction of Electrospun SiO2 Nanofibers. *Sci Rep* **5**, (2015).

7. Yang, Y. *et al.* A review on silicon nanowire-based anodes for next-generation high-performance lithium-ion batteries from a material-based perspective. *Sustain Energy Fuels* **4**, 1577–1594 (2020).

8. Jiang, B. *et al.* Dual Core–Shell Structured Si@SiO$_x$@C Nanocomposite Synthesized via a One-Step Pyrolysis Method as a Highly Stable Anode Material for Lithium-Ion Batteries. *ACS Appl Mater Interfaces* **8**, (2016).

9. Li, Y. *et al.* Growth of conformal graphene cages on micrometre-sized silicon particles as stable battery anodes. *Nat Energy* **1**, (2016).

10. Nava, G., Schwan, J., Boebinger, M. G., McDowell, M. T. & Mangolini, L. Silicon-Core–Carbon-Shell Nanoparticles for Lithium-Ion Batteries: Rational Comparison between Amorphous and Graphitic Carbon Coatings. *Nano Lett* **19**, 7236–7245 (2019).

11. Son, Y. *et al.* Quantification of Pseudocapacitive Contribution in Nanocage-Shaped Silicon–Carbon Composite Anode. *Adv Energy Mater* **9**, (2019).

12. Ko, M. *et al.* Scalable synthesis of silicon-nanolayer-embedded graphite for high-energy lithium-ion batteries. *Nat Energy* **1**, 16113 (2016).

13. Kwon, H. J. *et al.* Nano/Microstructured Silicon–Carbon Hybrid Composite Particles Fabricated with Corn Starch Biowaste as Anode Materials for Li-Ion Batteries. *Nano Lett* **20**, 625–635 (2020).

14. Wang, W. *et al.* Silicon and Carbon Nanocomposite Spheres with Enhanced Electrochemical Performance for Full Cell Lithium Ion Batteries. *Sci Rep* **7**, 44838 (2017).

15. Chae, S. *et al.* A Micrometer-Sized Silicon/Carbon Composite Anode Synthesized by Impregnation of Petroleum Pitch in Nanoporous Silicon. *Advanced Materials* **33**, 2103095 (2021).

16. Liu, N. *et al.* A pomegranate-inspired nanoscale design for large-volume-change lithium battery anodes. *Nat Nanotechnol* **9**, 187–192 (2014).

17. Heath, G. A. *et al.* Research and development priorities for silicon photovoltaic module recycling to support a circular economy. *Nat Energy* **5**, 502–510 (2020).
18. Jin, E. M. *et al.* Upcycling of silicon scrap collected from photovoltaic cell manufacturing process for lithium-ion batteries via transferred arc thermal plasma. *Energy* **262**, 125447 (2023).
19. Kasukabe, T. *et al.* Beads-Milling of Waste Si Sawdust into High-Performance Nanoflakes for Lithium-Ion Batteries. *Sci Rep* **7**, 42734 (2017).
20. Fan, Z. *et al.* Preparation of micron Si@C anodes for lithium ion battery by recycling the lamellar submicron silicon in the kerf slurry waste from photovoltaic industry. *Diam Relat Mater* **107**, 107898 (2020).
21. Lu, B. *et al.* Photovoltaic Monocrystalline Silicon Waste-Derived Hierarchical Silicon/Flake Graphite/Carbon Composite as Low-Cost and High-Capacity Anode for Lithium-Ion Batteries. *ChemistrySelect* **2**, 3479–3489 (2017).
22. Zhu, G., Wang, Y., Yang, S., Qu, Q. & Zheng, H. Correlation between the physical parameters and the electrochemical performance of a silicon anode in lithium-ion batteries. *Journal of Materiomics* **5**, 164–175 (2019).
23. Berhaut, C. L. *et al.* Prelithiation of silicon/graphite composite anodes: Benefits and mechanisms for long-lasting Li-Ion batteries. *Energy Storage Mater* **29**, 190–197 (2020).
24. Holtstiege, F., Bärmann, P., Nölle, R., Winter, M. & Placke, T. Pre-Lithiation Strategies for Rechargeable Energy Storage Technologies: Concepts, Promises and Challenges. *Batteries* **4**, 4 (2018).
25. Wang, F. *et al.* Prelithiation: A Crucial Strategy for Boosting the Practical Application of Next-Generation Lithium Ion Battery. *ACS Nano* **15**, 2197–2218 (2021).
26. McShane, E. J. *et al.* Quantification of Inactive Lithium and Solid–Electrolyte Interphase Species on Graphite Electrodes after Fast Charging. *ACS Energy Lett* **5**, 2045–2051 (2020).
27. Campbell, B. *et al.* Carbon-Coated, Diatomite-Derived Nanosilicon as a High Rate Capable Li-ion Battery Anode. *Sci Rep* **6**, 33050 (2016).
28. Hou, T. *et al.* The influence of FEC on the solvation structure and reduction reaction of LiPF6/EC electrolytes and its implication for solid electrolyte interphase formation. *Nano Energy* **64**, 103881 (2019).

29. Schroder, K. *et al.* The Effect of Fluoroethylene Carbonate as an Additive on the Solid Electrolyte Interphase on Silicon Lithium-Ion Electrodes. *Chemistry of Materials* **27**, 5531–5542 (2015).

30. Chen, S. *et al.* Constructing a Robust Solid–Electrolyte Interphase Layer via Chemical Prelithiation for High-Performance SiO$_x$ Anode. *Advanced Energy and Sustainability Research* **3**, 2200083 (2022).

31. Sun, Y. *et al.* In Situ Artificial Hybrid SEI Layer Enabled High-Performance Prelithiated SiO$_x$ Anode for Lithium-Ion Batteries. *Adv Funct Mater* (2023) doi:10.1002/adfm.202303020.

32. Su, Y.-S., Hsiao, K.-C., Sireesha, P. & Huang, J.-Y. Lithium Silicates in Anode Materials for Li-Ion and Li Metal Batteries. *Batteries* **8**, 2 (2022).

33. Raistrick, I. D., Ho, C. & Huggins, R. A. Ionic Conductivity of Some Lithium Silicates and Aluminosilicates. *J Electrochem Soc* **123**, 1469–1476 (1976).

34. Zhao, W. *et al.* Toward a stable solid-electrolyte-interfaces on nickel-rich cathodes: LiPO$_2$F$_2$ salt-type additive and its working mechanism for LiNi$_{0.5}$Mn$_{0.25}$Co$_{0.25}$O$_2$ cathodes. *J Power Sources* **380**, 149–157 (2018).

35. Ma, L. *et al.* LiPO$_2$F$_2$ as an Electrolyte Additive in Li[Ni$_{0.5}$Mn$_{0.3}$Co$_{0.2}$]O$_2$/Graphite Pouch Cells. *J Electrochem Soc* **165**, A891–A899 (2018).

36. Li, C. *et al.* Lithium Difluorophosphate as an Effective Additive for Improving the Initial Coulombic Efficiency of a Silicon Anode. *ChemElectroChem* **7**, 3743–3751 (2020).

37. Li, T. *et al.* Degradation Mechanisms and Mitigation Strategies of Nickel-Rich NMC-Based Lithium-Ion Batteries. *Electrochemical Energy Reviews* **3**, 43–80 (2020).

# Supporting Information

## Recycling Silicon Scrap for Spherical Si-C composite as High-Performance Lithium-ion Battery Anodes

*Bhagath Sreenarayanan[a], Marta Vicencio[a], Shuang Bai[b], Bingyu Lu[a], Ou Mao[d], Shiva Adireddy[d], Wurigumula Bao[c], and Ying Shirley Meng[a,b,c]\**

[a]Department of NanoEngineering, University of California San Diego, La Jolla, CA 92093

[b]Materials Science and Engineering, University of California San Diego, La Jolla, CA 92093

[c]Pritzker School of Molecular Engineering, University of Chicago, Chicago, IL 60637

[d]ADVANO, Inc. 2045 Lakeshore Drive, New Orleans, LA 70122

*Correspondence to wubao@uchicago.edu; shirleymeng@uchicago.edu

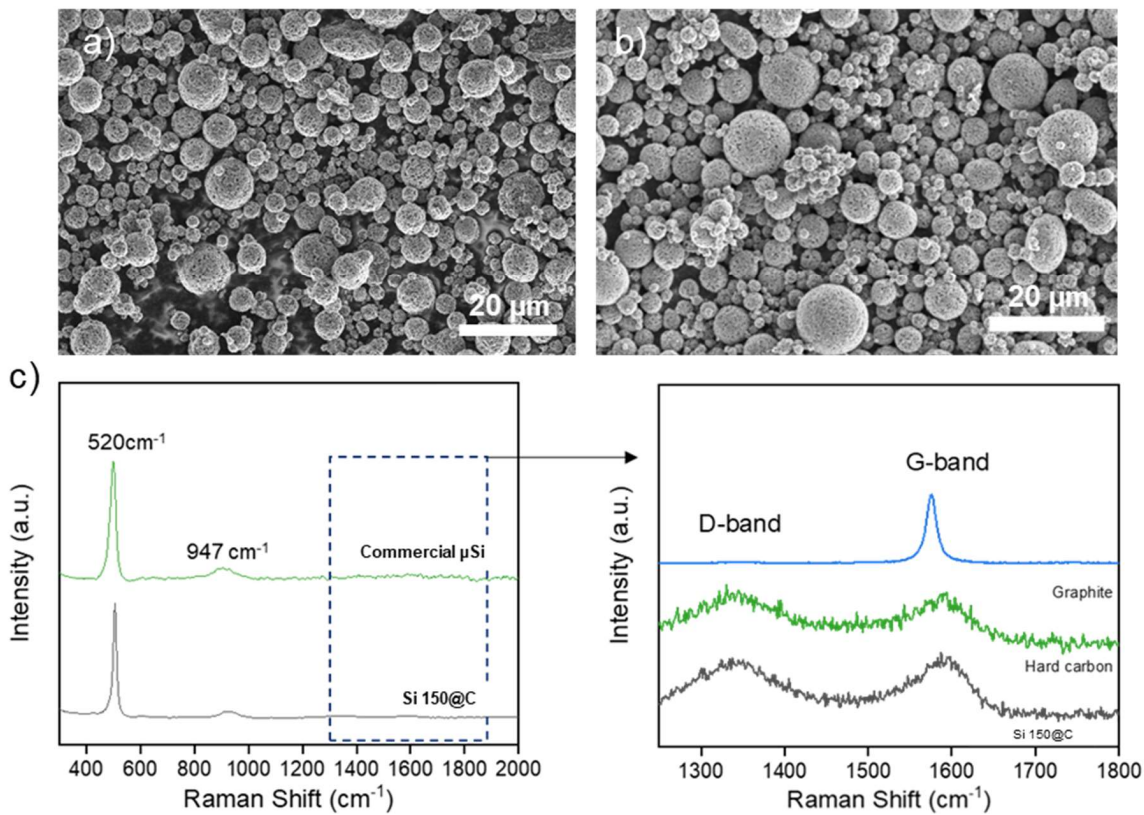

**Figure S1** Secondary Particle Distribution of (a) Si 150@C powder (b) S 80@C powder (c) Raman Spectra of Si 150@C, commercial µSi powder, Hard Carbon and Graphite

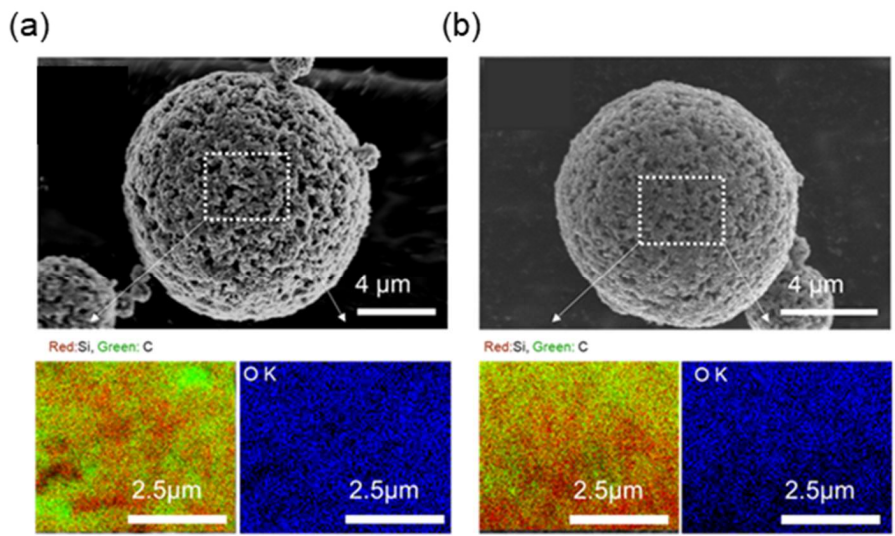

**Figure S2** SEM and EDS of secondary particle of (a) Si 150@C (b) Si 80@C

**Note S1**

Oxygen content % = (Total weight after TGA) - (Total Weight gain%=amount of $SiO_2$ formed)- Total weight after TGA/ 60*28)

Applying the above relation, from the TGA/DSC data (**From Figure 2 (a and b)**), we can see that:

For Si 150@C, Oxygen content % = 99 - 46.75- (99/60*28) =6%

For Si 80@C, Oxygen content% = 86-33.7-(86/60*28) =12%

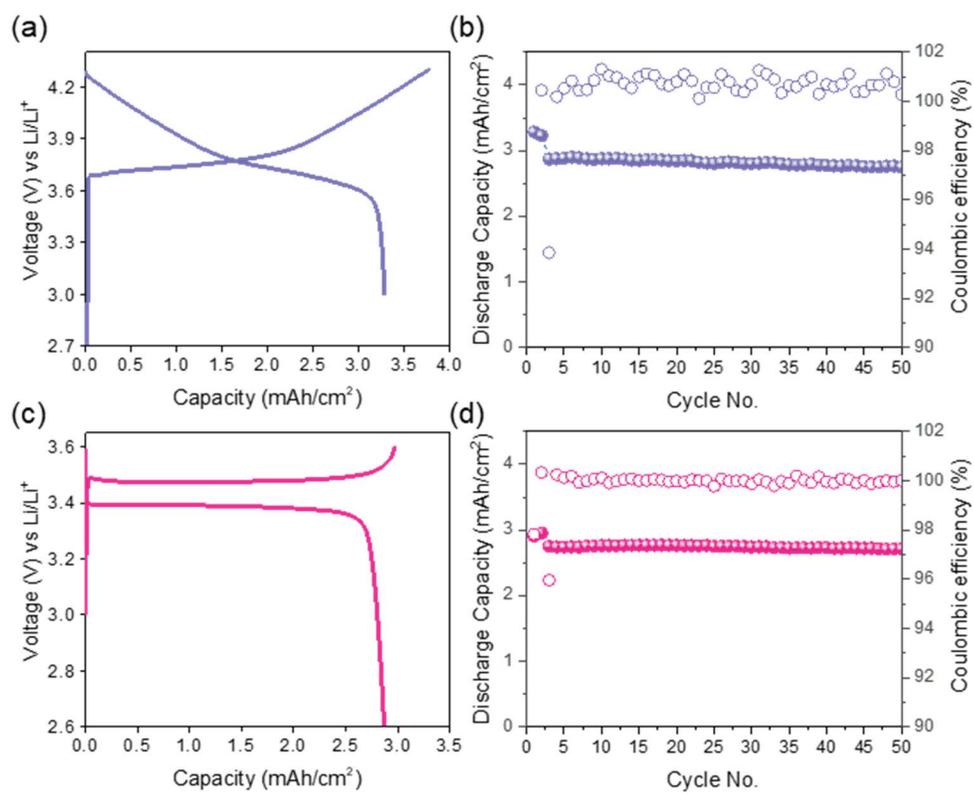

**Figure S3** (a) First Charge-Discharge curve for NMC622 in half cell (b) following cycle performance (c) First Charge-Discharge curve for LFP in half cell (b) following cycle performance

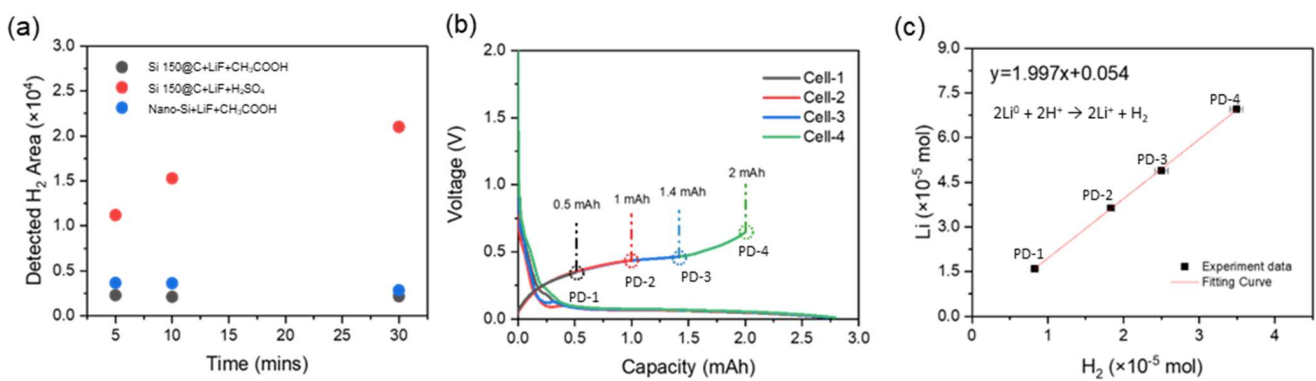

**Figure S4** (a) Stability check of Si 150@C and Nano Si+LiF samples with Acetic acid and $H_2SO_4$ (b) first cycle voltage profile of Si 150@C half cells with various de-lithiated states (PD-X, X=1,2,3..) (c) Calibration curve for acetic acid as titrant for TGC study.

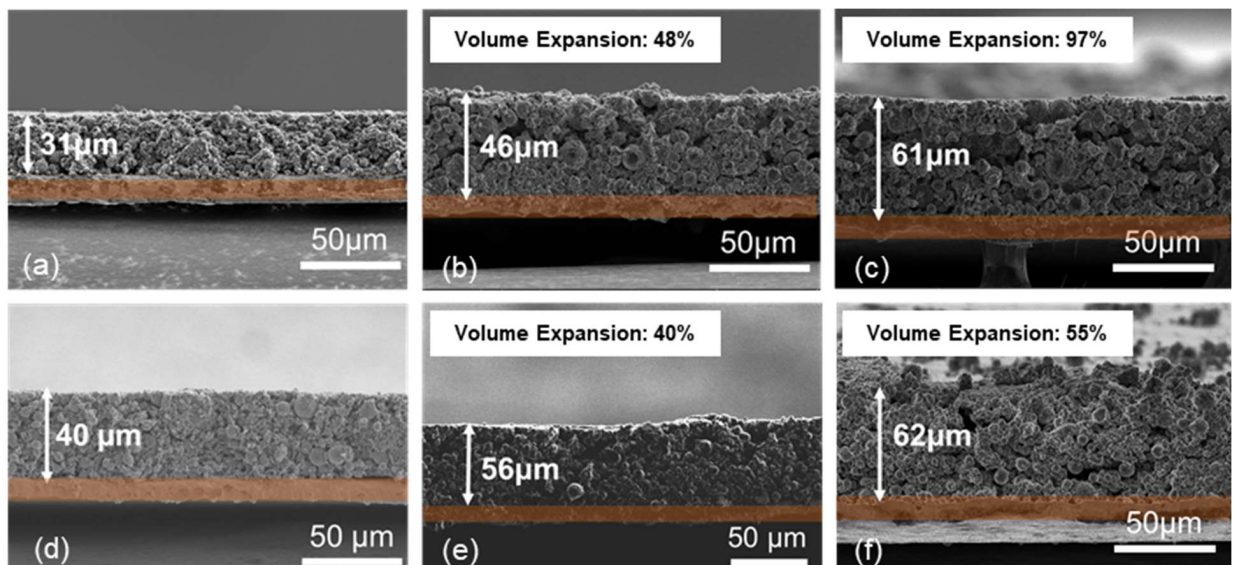

**Figure S5** SEM cross section images of Si 150@C electrode at (a) pristine, (b) 3<sup>rd</sup> cycle and (c) 50<sup>th</sup> cycle in full cell. SEM cross section images of Si 80@C electrode at (d) pristine, (e) 3<sup>rd</sup> cycle and (f) 50<sup>th</sup> cycle in full cell.

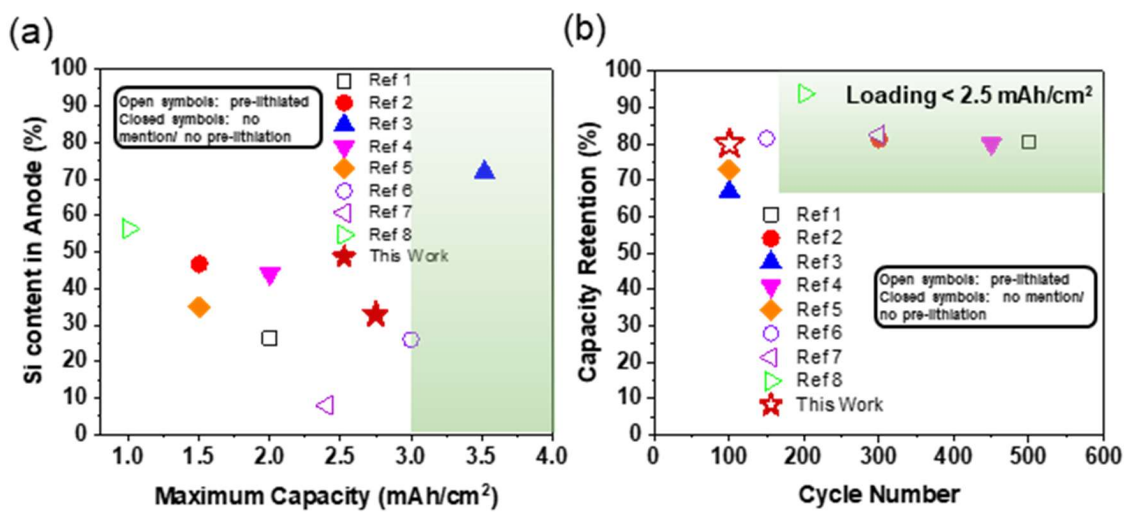

**Figure S6** (a) Si content in the anode (%) vs Maximum Capacity (mAh/cm$^2$) reported in the literature[1–8] (b) Capacity Retention (%) vs Cycle Number reported in the literature [1–8]

## Supplemental References

1. Kwon, H. J. *et al.* Nano/Microstructured Silicon–Carbon Hybrid Composite Particles Fabricated with Corn Starch Biowaste as Anode Materials for Li-Ion Batteries. *Nano Lett* **20**, 625–635 (2020).

2. Kobayashi, N., Inden, Y. & Endo, M. Silicon/soft-carbon nanohybrid material with low expansion for high capacity and long cycle life lithium-ion battery. *J Power Sources* **326**, 235–241 (2016).

3. Wang, W. *et al.* Silicon and Carbon Nanocomposite Spheres with Enhanced Electrochemical Performance for Full Cell Lithium Ion Batteries. *Sci Rep* **7**, 44838 (2017).

4. Chae, S. *et al.* A Micrometer-Sized Silicon/Carbon Composite Anode Synthesized by Impregnation of Petroleum Pitch in Nanoporous Silicon. *Advanced Materials* **33**, 2103095 (2021).

5. Son, Y. *et al.* Quantification of Pseudocapacitive Contribution in Nanocage-Shaped Silicon–Carbon Composite Anode. *Adv Energy Mater* **9**, 1803480 (2019).

6. Ma, Q. *et al.* Converting micro-sized kerf-loss silicon waste to high-performance hollow-structured silicon/carbon composite anodes for lithium-ion batteries. *Sustain Energy Fuels* **4**, 4780–4788 (2020).

7. Li, P., Hwang, J.-Y. & Sun, Y.-K. Nano/Microstructured Silicon–Graphite Composite Anode for High-Energy-Density Li-Ion Battery. *ACS Nano* acsnano.9b00169 (2019) doi:10.1021/acsnano.9b00169.

8. Chen, S., Shen, L., van Aken, P. A., Maier, J. & Yu, Y. Dual-Functionalized Double Carbon Shells Coated Silicon Nanoparticles for High Performance Lithium-Ion Batteries. *Advanced Materials* **29**, 1605650 (2017).

**Supplementary Energy Density Calculations**

**Assumptions:** Pouch Cell Type, package and tab mass excluded; Pre-lithiation is not taken into consideration for Si 80@C-LFP cell

**Model:** Anode|Liquid Electrolyte|Cathode, Double Coating

a) For Si 150@C-NMC622 Pouch Cell

| Parameters for Calculation | Value | Remarks |
|---|---|---|
| Nominal Voltage (V) | 3.4 | Average Plateau of Voltage Curve |
| Capacity (Ah) | 294.1176 | For 1 kWh |
| Utilization Rate (Fraction) | 1 | |
| Areal Capacity (mAh/cm$^2$) | 2.76 | Taken as Discharge Capacity from Full Cell Data |
| Total Area (cm$^2$) | 106564.4 | |
| N/P ratio (Fraction) | 1.1 | |
| Liquid Electrolyte Ratio (g/Ah) | 1 | |
| Liquid Electrolyte Density (g/cc) | ~1.3 | |
| Total Electrolyte Required (g) | 294.1176 | |
| Separator Mass (mg/cm$^2$) | 1.386 | Celgard 2325 |
| Total Separator Mass (g) | 147.6982 | Minimal Area to cover the cathode |
| Cathode Active Ratio (Fraction) | 0.9 | |
| Cathode Lithiation Capacity (mAh/g) | 180 | Discharge Capacity |
| Total Cathode Required (g) | 1633.987 | |
| Aluminum Thickness (cm) | 0.0016 | Commercial Standards |
| Aluminum Density (g/cc) | 2.7 | |
| Total Aluminum Required (g) | 230.179 | |
| Binder/Carbon ratio in Cathode (Fraction) | 0.05 | |
| Total Binder and Carbon Required (g) | 90.77705 | |
| Anode Active Ratio (Fraction) | 0.7 | |
| Anode De-Lithiation Capacity (mAh/g) | 1530 | |
| Total Anode Required (g) | 211.4571 | |
| Copper Thickness (cm) | 0.0009 | Commercial Standards |
| Copper Density (g/cc) | 8.96 | |
| Total Copper Required (g) | 429.6675 | |
| Binder/Carbon ratio in Anode (Fraction) | 0.15 | |
| Total Binder and Carbon Required (g) | 45.31224 | |
| Total Mass (kg) | 3.08319 | |
| Energy Density at Cell Level (Wh/kg) | **324.3388** | |

**Table S1** Energy Density Calculation for Si 150@C-NMC622 Pouch type Cell

b) For Si 80@C-LFP Pouch Cell

| Parameters for Calculation | Value | Remarks |
|---|---|---|
| Nominal Voltage (V) | 3.1 | Average Plateau of Voltage Curve |
| Capacity (Ah) | 322.5806 | For 1 kWh |
| Utilization Rate (Fraction) | 1 | |
| Areal Capacity (mAh/cm$^2$) | 2.25 | Taken as Discharge Capacity from Full Cell Data |
| Total Area (cm$^2$) | 143369.2 | |
| N/P ratio (Fraction) | 1.1 | |
| Liquid Electrolyte Ratio (g/Ah) | 1 | |
| Liquid Electrolyte Density (g/cc) | ~1.3 | |
| Total Electrolyte Required (g) | 322.5806 | |
| Separator Mass (mg/cm$^2$) | 1.386 | Celgard 2325 |
| Total Separator Mass (g) | 198.7097 | Minimal Area to cover the cathode |
| Cathode Active Ratio (Fraction) | 0.9 | |
| Cathode Lithiation Capacity (mAh/g) | 160 | Discharge Capacity |
| Total Cathode Required (g) | 2016.129 | |
| Aluminum Thickness (cm) | 0.0016 | Commercial Standards |
| Aluminum Density (g/cc) | 2.7 | |
| Total Aluminum Required (g) | 309.6774 | |
| Binder/Carbon ratio in Cathode (Fraction) | 0.05 | |
| Total Binder and Carbon Required (g) | 112.0072 | |
| Anode Active Ratio (Fraction) | 0.7 | |
| Anode De-Lithiation Capacity (mAh/g) | 1400 | |
| Total Anode Required (g) | 253.4562 | |
| Copper Thickness (cm) | 0.0009 | Commercial Standards |
| Copper Density (g/cc) | 8.96 | |
| Total Copper Required (g) | 578.0645 | |
| Binder/Carbon ratio in Anode (Fraction) | 0.15 | |
| Total Binder and Carbon Required (g) | 54.31205 | |
| Total Mass (kg) | 3.844937 | |
| Energy Density at Cell Level (Wh/kg) | **260.0823** | |

**Table S2** Energy Density Calculation for Si 80@C-LFP Pouch type Cell